# Characterization of free standing InAs quantum membranes by standing wave hard x-ray photoemission spectroscopy.

G. Conti[1,2], S. Nemšák[1,2,3,4], C.-T. Kuo[1,2], M. Gehlmann[1,2,3], C. Conlon[1,2], A. Keqi[1,2], A. Rattanachata[1,2], O. Karslıoğlu[5], J. Mueller[6], J. Sethian[11], H. Bluhm[4,5], J. E. Rault[7], J. P. Rueff[7,8], H. Fang[9], A. Javey[2,10], and C. S. Fadley[1,2]

[1]Department of Physics, University of California Davis, Davis, California 95616, USA

[2]Materials Sciences Division, Lawrence Berkeley National Laboratory, Berkeley, California 94720, USA

[3]Peter-Grünberg-Institut PGI-6, Forschungszentrum Jülich, 52425 Jülich, Germany, Europa

[4]Advanced Light Source, Lawrence Berkeley National Laboratory, Berkeley, California 94720, USA

[5]Chemical Sciences Division, Lawrence Berkeley National Laboratory, Berkeley, California 94720, USA

[6]Center for Computational Sciences and Engineering, Lawrence Berkeley National Laboratory, Berkeley, California 94720, USA

[7]Synchrotron-SOLEIL, BP 48, Saint-Aubin, F91192 Gif sur Yvette CEDEX, France, Europe

[8]Sorbonne Université, UPMC Univ Paris 06, CNRS, UMR 7614, Laboratoire de Chimie Physique - Matière et Rayonnement, 75005 Paris Cedex 05, France, Europe

[9]Department of Electrical and Computer Engineering, Northeastern University, Boston, Massachusetts 02115, USA

[10]Electrical Engineering and Computer Sciences, University of California Berkeley, Berkeley, California 94720, USA

[11]Department of Mathematics, University of California Berkeley, Berkeley, California 94720, USA

**ABSTACT:**

Free-standing nanoribbons of InAs quantum membranes (QMs) transferred onto a (Si/Mo) multilayer mirror substrate are characterized by hard x-ray photoemission spectroscopy (HXPS), and by standing-wave HXPS (SW-HXPS). Information on the chemical composition and on the chemical states of the elements within the nanoribbons was obtained by HXPS and on the quantitative depth profiles by SW-HXPS. By comparing the experimental SW-HXPS rocking curves to x-ray optical calculations, the chemical depth profile of the InAs(QM) and its interfaces were quantitatively derived with angstrom precision. We determined that: i) the exposure to air induced the formation of an $InAsO_4$ layer on top of the stoichiometric InAs(QM); ii) the top interface between the air-side $InAsO_4$ and the InAs(QM) is not sharp, indicating that interdiffusion occurs between these two layers; iii) the bottom interface between the InAs(QM) and the native oxide $SiO_2$ on top of the (Si/Mo) substrate is abrupt. In addition, the valence band offset (VBO) between the InAs(QM) and the $SiO_2$/(Si/Mo) substrate was determined by HXPS. The value of VBO = 0.2±0.04 eV is in good agreement with literature results obtained by electrical characterization, giving a clear indication of the formation of a well-defined and abrupt InAs/$SiO_2$ heterojunction. We have demonstrated that HXPS and SW-HXPS are non-destructive, powerful methods for characterizing interfaces and for providing chemical depth profiles of nanostructures, quantum membranes, and 2D layered materials.

III-V compound semiconductors possess superb carrier transport and excellent optoelectronic properties, which render them widely used in high performance electronic and optoelectronic devices, such as high electron mobility transistors, heterostructure lasers and solar cells.[1,2,3] These III-V "alternative" semiconductors, such as InAs and InGaSb, have much higher electron/hole mobility than Si and are good candidates to replace Si as future channel materials in metal–oxide–semiconductor field-effect transistors (MOSFETs).[4,5,6,7] However, the growth of these semiconductors and their integration with the low-cost processing of Si technology present challenges, since high defect densities and junction leakage currents may occur at the interface between these compounds and the Si substrate[8] due the their lattice mismatch. In order to overcome the problems related to the growth of these semiconductors on Si or $SiO_2$ substrates and to their integration with the more mature Si-based processes, Javey *et al.*[9] developed a method for transferring ultrathin free-standing crystalline III-Vs films on a user-defined substrate. This transfer step led to high-performance III-V complementary metal–oxide–semiconductor (CMOS) and radio frequency (RF) circuits on both Si and plastic substrates.[10] However, the non-destructive characterization of these transferred films is still a challenge. In particular, their depth-resolved chemical composition and the intermixing/oxidation occurring at the interfaces have not been fully determined. Low- and high- resolution transmission electron microscopy (TEM) characterization on such membranes has been performed by Javey *et al.*.[11] However, it is well known that the sample preparation for TEM analysis can induce damages to the material and can create interdiffusion at the interfaces.[12] Therefore, TEM has limitations as a technique for a non-destructive, quantitative characterization of the interfaces of quantum membranes. On the other hand, the characterization of these interfaces is fundamental for high performance nanoscale transistors. It is known that oxide layers on the surface[13], or at the interface, strongly influence the electronic transport properties of these materials, especially in the case of nanowires and quantum membranes. For instance, the native oxide layer could shift the position of the surface Fermi-level or even induce Fermi-level pinning, which significantly degrades device performances.[14,15,11]

In this paper, we show that hard x-ray photoemission spectroscopy (HXPS) and Standing Wave HXPS (SW-HXPS) can be used to characterize bulks and buried layers/buried interfaces without altering the samples. Standard XPS analysis provides information on the chemical and electronic states of each element at the surface. Moving to hard x-ray photon energies (2–15 keV), and thus to larger electron inelastic mean-free paths (IMFPs), allows one to obtain information on chemical and electronic states of the bulk and of the buried interfaces.[16] Among the many III-V alternative semiconductors, we chose to study free-standing InAs quantum membranes (InAs(QM)) which, in addition to their interesting properties, represent a more general class of 2-D materials. For instance, it has been reported[9,11] that when the thickness of InAs is reduced, especially when it is below its exciton Bohr radius (34 nm), strong quantum confinement effects start to emerge and its band structure can be precisely tuned from bulk to 2D by changing the thickness. In particular, we have studied by SW-HXPS the possible formation of a surface oxide layer and interdiffusion at the top and bottom interfaces of InAs(QM) nanoribbons. Furthermore, we have determined if and what kind of oxide layer is formed at the top vacuum/InAs interface and how abrupt the InAs/bottom layer interface is.

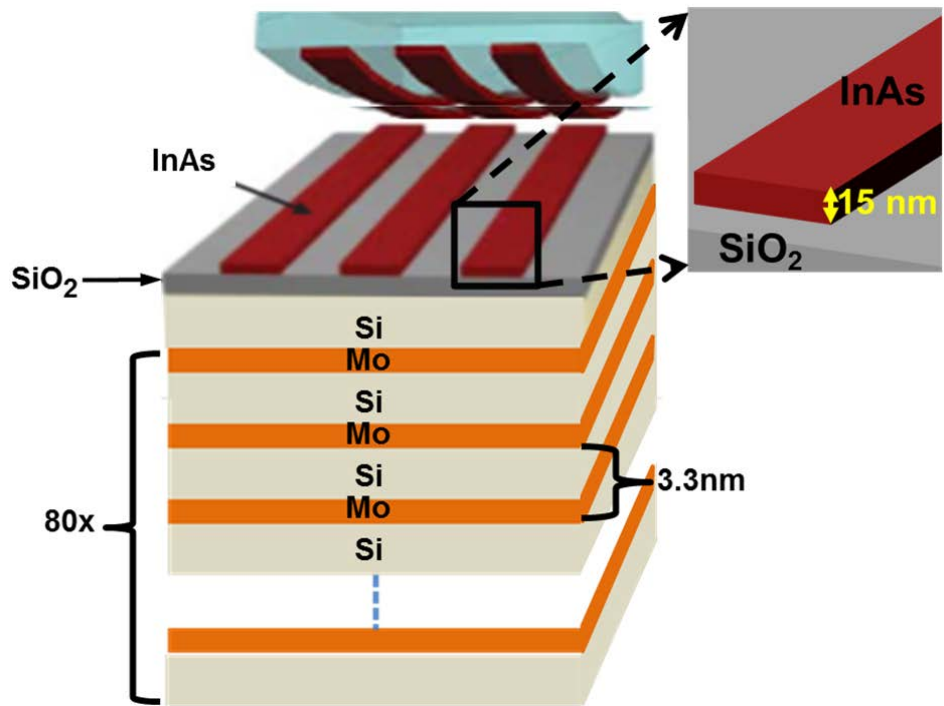

Figure 1: Schematic of the InAs(QM) nanoribbons transferred onto a (Si/Mo)x80 multilayer (dimensions not to scale). Each nanoribbon is 15 nm thick and 300 nm wide and the distance between one ribbon and its neighbors is 300 nm. The multilayer period is 3.3 nm.

In order to perform the SW-HXPS experiment, free-standing crystalline nanoribbons of InAs(QM) were transferred onto a (Si/Mo)x80 periodic multilayer mirror with the bilayer period of $d_{ML}=$ 3.3 nm (see Figure 1), which generates a standing wave by reflecting the x-rays at the incidence angle defined by the first-order Bragg reflection[17]

$$\lambda_x = 2d_{ML}\sin(\theta_{inc}) \qquad (1)$$

where $\lambda_x$ is the incident photon wavelength, $d_{ML}$ is the period of the multilayer mirror, and $\theta_{inc}$ is the incidence angle. The wavelength of the standing wave so generated is

$$\lambda_{SW}(|E^2|) = \frac{\lambda_x}{2\sin\theta_{inc}} = d_{ML} \qquad (2)$$

and matches the period of the multilayer mirror. For a given photon energy, by varying the angle $\theta_{inc}$ around the Bragg reflection, the phase of the standing wave varies over π. As the antinodes of the electromagnetic field shift vertically through the sample, they highlight different depths in the sample. This provides depth selectivity to the photoemission process. In our case, the standing wave, which travels perpendicularly to the multilayer and to the sample surfaces deposited on it, allows us to obtain a quantitative chemical depth profile of the InAs(QM) and of its interfaces with vacuum and with the SiO$_2$/mirror substrate. The vertical resolution is approximately 1/10 of the SW period, which is ≈ 0.3 nm for the mirror used in this study.[18,19]

The InAs(QM) in the shape of nanoribbons was epitaxially grown on the [111] plane. Each nanoribbon is 15 nm thick and 300 nm wide and the distance between one ribbon and its neighbors is 300 nm (Figure 1). The (Si/Mo) multilayer onto which these InAs(QM) nanoribbons were transferred, was prepared at the Center for X-ray Optics of the Lawrence Berkeley National Laboratory, and consists of 80 (Si/Mo) bilayers, each bilayer having a thickness of 3.3 nm. The termination layer of this (Si/Mo) mirror was chosen to be Si, which, exposed to air, gives rise to a thin layer of native silicon oxide (SiO$_2$). In this way, the bottom interface between the InAs(QM) and the SiO$_2$/(Si/Mo) substrate is a good approximation to the interface between the InAs(QM) and a typical silicon-based substrate. Before transferring the InAs(QM) on it, the mirror was cleaned with acetone, isopropyl alcohol, and de-ionized water.[9] At the photon energy of 4.0 keV used in our measurements, the Bragg angle is $\theta_{Bragg}$ ≈ 2.7°. In order to scan over the first order Bragg reflection, the incidence angle of the incoming x-ray beam was varied between 2.2° and 3.8°, in steps 0.02°. The HXPS spectra were obtained at the Advanced Light Source (ALS) (Beamline 9.3.1) and at SOLEIL Synchrotron (GALAXIES Beamline).[20] The p-polarized x-ray photon energy was set to hv = 4.0 keV and the spectral total energy resolution was ≈ 500 meV for the data acquired at ALS and ≈ 250 meV for the data acquired at SOLEIL. At hv = 4.0 keV, the IMFP, as estimated from the TPP-2M formula,[21] is ~8 nm for InAs. As a consequence, not only the top surface, but also the bottom interface between the 15 nm thick InAs(QM) and the SiO$_2$(Si/Mo) substrate could be characterized by HXPS. The binding energies of the HXPS spectra were calibrated using Au 4f and Au E$_F$ before and after each data acquisition.

Figure 2 shows the relevant core level (CL) spectra of the InAs(QM) and of the SiO$_2$/(Si/Mo) mirror. Various chemically-shifted components are also indicated. Since the InAs(QM) sample was exposed to air before the HXPS analysis, the presence of C 1s is probably due to surface contamination. Although C 1s was fit with two components, their rocking curves (RCs) are identical, thus, there is only one effective depth for whatever C contaminant species are present.

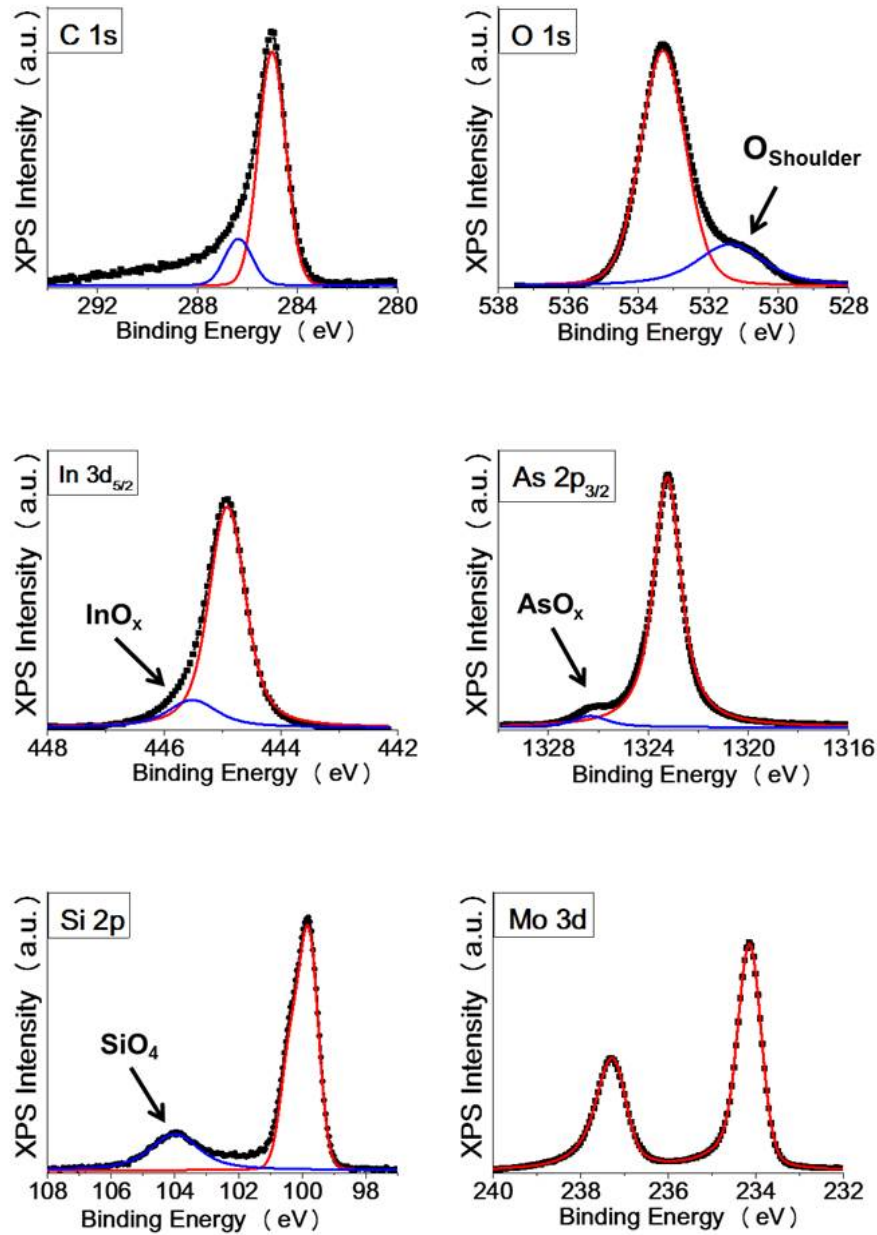

Figure 2: Experimental x-ray photoelectron core level spectra for all of the elements in the InAs(QM) and the mirror substrate: C 1s, O 1s, As $2p_{3/2}$, In $3d_{5/2}$, Mo 3d and Si 2p, with chemically-shifted components indicated.

Each of the As 2p, In 3d, Si 2p, and O 1s CL spectra shows clear evidence of two components. As reported in the literature[22,23], the As 2p shoulder at binding energy (BE) 1326 eV and that of In 3d at BE=446 eV are attributed to the presence of AsO$_x$ and InO$_x$, and the weak Si 2p peak at BE=103.5 eV is assigned to the native SiO$_2$ on top of the (Si/Mo) mirror. The depth distribution of these elements and of their different valence states can be obtained by SW-HXPS, including the two O 1s components, the main peak at BE=533.4 eV and the shoulder at BE=531 eV. The CL spectra were analyzed by subtracting a Shirley inelastic background[24] from each spectrum before fitting it with Voigt functions. The peak integrated intensity of each CL was then plotted versus the incidence angle, thus generating what we call experimental rocking curves (RCs). The RCs so obtained are shown in figure 3 (data points). From these RCs a qualitative picture of the chemical depth profile of the InAs(QM) nanoribbons can be obtained. All the RCs show a maximum or minimum at the incidence angle ≈ 2.7° which corresponds to the Bragg peak of the (Si/Mo) multilayer. In addition we observe that the RCs can be grouped into four different categories:1) the RC of C 1s stands alone and its phase does not match with any others, as might be expected of a species with a single surface location; 2) the RCs of As 2p$_{3/2}$ shoulder at BE=1326 eV, of In 3d$_{5/2}$ shoulder at BE=446 eV and of O1s shoulder at BE=531 eV are very similar in phase and intensity modulation, indicating that these In, As and O atoms are at the same depth and suggesting the presence of an InAsO$_x$ layer; 3) the RCs of As 2p$_{3/2}$ main peak at BE=1323 eV and of In3d$_{5/2}$ main peak at BE=445 eV are very similar in phase and intensity modulation, although their phase is opposite and their modulation is smaller than those of the RCs in group 2. This indicates the presence of an InAs layer thicker and at a different depth than the InAsO$_x$ layer. The lower modulation of the InAs layer is due to the large escape depth of the electrons, coupled with the high layer thickness of the InAs compared to the SW period, with the thickness extending over multiple periods of the SW; 4) the RCs of the O 1s shoulder at BE=531 eV and of Si 2p at BE=104 eV show the same modulation and phase, confirming that the O 1s shoulder is associated with the presence of the native SiO$_2$ on top of the (Si/Mo) mirror.

In addition to this qualitative information, an analysis of the experimental RCs using an x-ray optical program, the Yang X-Ray Optics (YXRO) code[17,25], can provide more quantitative information on the depth, chemical state, composition and interdiffusion of the four layers described above. In order to accurately predict the RCs, the YXRO code, in addition to the sample optical parameters, takes into account the differential photoelectric cross sections and the photoelectron IMFPs. The calculated RCs reported in this paper were obtained by combining the YXRO code with a global black box optimizer[26] in order to maximize the goodness of the fit and the speed of the fitting process. A conventional squared-deviation R-factor was used to judge the goodness of the fit as a function of the sample structural parameters: layer thicknesses and interface thicknesses. This combination of the two programs has led to an approximately 100-fold increase in the speed of analysis compared to prior hand fitting.[26]

Figure 3 shows the final results for the experimental RCs (data points) and calculated RCs (curves) obtained by YXRO combined with the optimizer. The comparison between the experimental and calculated RCs for the final best-fit parameters thus provides quantitative information on the constituent layers and on the interfaces, including mixing or roughness at the interface.

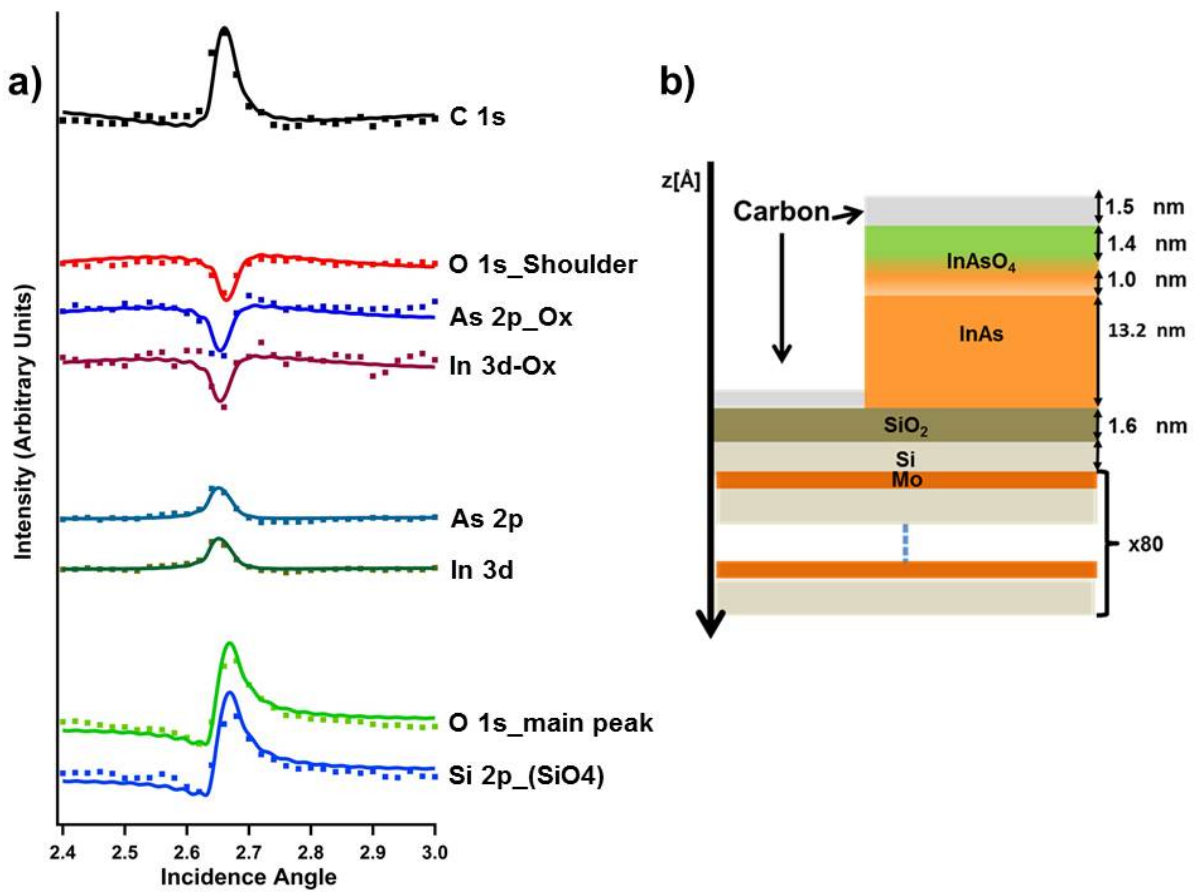

Figure 3: a) Experimental (data points) and calculated (curves) RCs. b) Summary of InAs(QM) nanoribbons depth profile.

The agreement between the experimental RCs and the simulations from our best-fit is remarkable (Figure 3a). The results from the optimized depth profile are reported in Figure 3b. The top layer of adventitious carbon is ≈ 1.5 nm thick, in good agreement with the C 1s core level relative intensity as simulated by the surface analysis program SESSA.[27] The C 1s RC is special, in that there are two contributors, on top of the InAs(QM) and on top of the $SiO_2$, but the vertical distance (≈ 17.1 nm) separating them is about (1.4+0.9+13.2+0.8)= 16.3 nm from the top layer and 0.8 nm from the bottom layer, and this is almost 5 periods of the SW = 16.6 nm. Thus, the intensities of these two layers add coherently in the RC. The top layer of InAs(QM) is oxidized. The exposure to air of the InAs(QM) nanoribbons induced the formation of a layer of what is probably InAsO$_4$[28,22] which is ≈ 1.4 nm thick. The remaining part of the nanoribbons consists of a stoichiometric InAs layer of ≈ 13.2 nm. The top interface between the InAsO$_4$ and the InAs layers is not sharp but has ≈ 0.9 nm interdiffusion. On the other hand, the bottom interface between the InAs(QM) and the SiO$_2$/(Si/Mo) mirror is abrupt and the thickness of the native SiO$_2$ on top of the mirror is ≈ 1.6 nm.

Having determined that there is negligible interdiffusion between the bottom InAs(QM) and the SiO$_2$/mirror substrate, the valence band offset (VBO) between these InAs(QM) nanoribbons and the SiO$_2$/(Si/Mo) substrate could be determined by HXPS, using the method originally introduced by Kraut *et al.*.[29,30]

The VBO is the key parameter for designing high performance nanoscale transistors and electronic/photonic devices etc., therefore the determination of this parameter is critically important. So far, only a limited amount of experimental and theoretical information regarding band alignment at the interfaces of nanoparticles or quantum membranes is available.[31,32] Experimentally, the most direct and reliable way for determining the band alignment is by using core levels and valence band maximum binding energies from XPS, as originally demonstrated by Kraut *et al.*.[29,30] This method has been widely used in semiconductor applications, even though it does not consider differences in screening of core and valence holes near an interface compared to the reference bulk materials. Chambers *et al.*[33] *and references therein*, explored the influence on the measurement of VBO by XPS of the core-level peak broadening due to chemical effects and electronic charge redistributions at the surface and interface for thin-film heterojunctions. In their studies of semiconductor and transition-metal heterostructures, this group has noticed that core-level binding energies of epitaxial interfaces of metals on oxides, and oxides on metals do not change in a significant way as a function of film thickness, implying that differences in screening of core and valence holes near an interface compared to pure material do not significantly impact the determination of VBO by using the Kraut method, provided that chemical effects and electronic charge redistributions are not involved. We therefore believe that it is appropriate to determine the VBO between the InAs(QM) nanoribbons transferred on the SiO$_2$/(Si/Mo) substrate using this method, which says that

$$\Delta E_{VBO}(A/B) = (E_{CL}^{A'} - E_{CL}^{B'}) - [(E_{CL}^{A} - E_{VBM}^{A}) - (E_{CL}^{B} - E_{VBM}^{B})] \tag{3}$$

where $\Delta E_{VBO}$ is the VBO of layer A (InAs) relative to layer B (mirror substrate) , $E_{CL}^{A'orB'}$ is the binding energy of a core level A' or B' in the InAs/Mirror heterostructure; $E_{CL}^{A,orB}$, and $E_{VBM}^{A,orB}$ are the binding energy of the same core levels and the valence band maximum in the two bulk materials InAs and SiO$_2$/(Si/Mo) mirror, respectively.

Thus, we acquired HXPS core levels and valence band spectra of bulk InAs [111] and of a stand-alone SiO$_2$/(Si/Mo) mirror, and the core levels of the InAs(QM)[111] nanoribbons on the mirror. HXPS spectra did not show any broadening of the core levels in our sample compared to the bulk references. Using equation (3) we determined that the VBO of InAs(QM) relative to SiO$_2$(1.6nm)/(Si/Mo) substrate is *0.2±0.04 eV.* This value is in good agreement with literature results, which span ~0.1-0.2 eV,[34,35,36,37] as obtained by electrical characterization, especially considering the differences between our InAs(QM)/SiO$_2$ interface and the nanostructures previously studied in the literature.

In conclusion, we have demonstrated that HXPS and SW-HXPS are powerful non-destructive methods that can be used to characterize the interfaces in novel materials and nanostructures such as quantum membranes. For instance, HXSP and SW-HXPS provided the stoichiometry, the depth and the thickness of the oxide overlayer on these InAs(QM) nanoribbons, which acts as passivation layer, useful in order to prevent dangling bonds. In addition, SW-HXPS showed that the interface between the InAs(QM) and its oxidation layer is not sharp, indicating that some interdiffusion occurred and that the oxidation is not entirely homogenous. On the contrary, the bottom interface between the InAs(QM) and the substrate is atomically abrupt, which is a crucial prerequisite for successful applications of high performance nanoscale transistors. In addition, the VBO between these InAs(QM) nanoribbons and the SiO2/(Si/Mo) substrate was determined. The obtained value of 0.2±0.04eV is in good agreement with literature results giving a clear indication of the formation of a well-defined and abrupt SiO$_2$–InAs heterojunction.

**Acknowledgments**

This work was supported by the US Department of Energy under Contract No. DE-AC02-05CH11231 (Advanced Light Source, Materials Sciences and Chemical Sciences Divisions), and by DOE Contract No. DE-SC0014697 through the University of California Davis (salary for C.-T.K, M.G. and C.S.F.). C.S.F. has also been supported for salary by the Director, Office of Science, Office of Basic Energy Sciences (BSE), Materials Sciences and Engineering (MSE) Division, of the U.S. Department of Energy under Contract No. DE-AC02-05CH11231, through the Laboratory Directed Research and Development Program of Lawrence Berkeley National Laboratory, through the APTCOM Project, "Laboratoire d'Excellence Physics Atom Light Matter" (LabEx PALM) overseen by the French National Research Agency (ANR) as part of the "Investissements d'Avenir" program, and from the Jűlich Research Center, Peter Grűnberg Institute, PGI-6. A.R. was funded by the Royal Thai Government, A.K was awarded with graduate student researcher-work study and departmental fellowship for at UC Davis, and C.C. was funded by GAANN program through UC Davis Physics Department. Materials processing was supported by the Electronic Materials Program funded by the Director, Office of Science, Office of Basic Energy Sciences, Materials Sciences and Engineering Division of the U.S. Department of Energy, under contract no. DE-AC02 05Ch11231. O.K. and H.B. acknowledge support by the Director, Office of Science, Office of Basic Energy Sciences (BSE), Chemicals Sciences, Geosciences and Biosciences, of the U.S. Department of Energy under Contract No. DE-AC02 05CH1123.

---